\documentclass[a4paper, oneside, twocolumn, notitlepage, 10pt]{extarticle_ecoc}
\usepackage{ecoc}
\usepackage{acronym}
\usepackage{multirow}
\usepackage{enumerate}
\acrodef{OLS}{Open Line System}
\acrodef{ROADM}{Reconfigurable Optical Add-Drop Multiplexer}
\acrodef{BVT}{Bandwidth Variable Transceiver}
\acrodef{OSNR}{Optical Signal-to-Noise Ratio}
\acrodef{ASE}{Amplified Spontaneous Emission}
\acrodef{NETCONF}{Network Configuration}
\acrodef{NLI}{Non-Linear Interference}
\acrodef{BER}{Bit Error Rate}
\acrodef{OSaaS}{Optical Spectrum as a Service}
\acrodef{CFO}{Carrier Frequency Offset}
\acrodef{CDC}{Chromatic Dispersion Compensation}
\acrodef{DGD}{Differential Group Delay}
\acrodef{ORP}{Optical Received Power}
\acrodef{SNR}{Signal to Noise Ratio}
\acrodef{PDL}{Polarization Dependent Loss}
\acrodef{ML}{Machine Learning}
\acrodef{ANN}{Artificial Neural Network}
\acrodef{CNN}{Convolutional Neural Network}
\acrodef{OCM}{Optical Channel Monitor}
\acrodef{OPM}{Optical Performance Monitoring}
\acrodef{EDFA}{Erbium Doped Fiber Amplifier}
\acrodef{EON}{Elastic Optical Network}
\acrodef{DWDM}{Dense Wavelength Division Multiplexing}
\acrodef{QoT}{Quality of Transmission}
\acrodef{MLP}{Multi Layer Perception}
\acrodef{SVM}{Support Vector Machines}
\acrodef{AGC}{Automatic Gain Control}
\acrodef{WSS}{Wavelength Selective Switch}
\acrodef{ReLU}{Rectified Linear Unit}
\usepackage{soul}
\begin{document}
\selectlanguage{english}

\title{Interference Identification in Multi-User Optical Spectrum as a Service using Convolutional Neural Networks}%
\vspace{-3mm}
\author{
    Agastya Raj\textsuperscript{*,}\textsuperscript{(1)}, 
    Zehao Wang\textsuperscript{(2)},
    Frank Slyne\textsuperscript{(1)},
    Tingjun Chen\textsuperscript{(2)}, 
    Dan Kilper\textsuperscript{(1)}, 
    Marco Ruffini\textsuperscript{(1)}
}
\vspace{-7mm}
\maketitle 
\begin{strip}
    \begin{author_descr}

        \textsuperscript{(1)} CONNECT Centre, School of Computer Science and Statistics and School of Engineering, Trinity
College Dublin, Ireland
        \textcolor{blue}{\uline{*rajag@tcd.ie}}

        \textsuperscript{(2)} Duke University, Department of Electrical and Computer Engineering, Durham, NC, USA

    \end{author_descr}
    \vspace{-3.5mm}
\end{strip}

\renewcommand\footnotemark{}
\renewcommand\footnoterule{}

\begin{strip}
    \begin{ecoc_abstract}
        We introduce a ML-based architecture for network operators to detect impairments from specific OSaaS users while blind to the users' internal spectrum details. Experimental studies with three OSaaS users demonstrate the model's capability to accurately classify the source of impairments, achieving classification accuracy of 94.2\%.  \textcopyright2024 The Author(s)
    \end{ecoc_abstract}
    \vspace{-4mm}
\end{strip}

\section{Introduction}
\vspace{-2mm}
The recent move of optical networks to 
\acp{EON} 
has enabled new flexible transport services, such as \ac{OSaaS}~\cite{OSaaS_JOCN, OSaaS_ECOC}. \ac{OSaaS} can be applied to systems in which network users can operate multiple optical channels (using their own transceivers), leasing a given spectral window in a third party fiber link (e.g., 400 GHz). 
This approach has been used commercially in \ac{OSaaS}~\cite{geant}, serving trusted entities where the user is another network operator.
Extending \ac{OSaaS} to more user types will create new fiber leasing models, making more efficient use of excess capacity. At the same time, it benefits users who could gain access to optical network spectrum without the costs and complications of managing an entire fiber. An example is that of mobile operators looking to operate an Open RAN fronthaul network across a metro area 
~\cite{wypior_open_2022}.

However, providing shared fiber access to third parties opens the network to many vulnerabilities. Due to inter-channel interference such as fiber non-linearity based crosstalk~\cite{pointurier_design_2017}, the dynamics of the channels within a given spectral window can affect the performance of other channels in the fiber link. 
Any user-initiated changes such as power adjustments or channel addition, can inadvertently introduce impairments for the operator's own channels, and to other \ac{OSaaS} users.   

One attraction of the \ac{OSaaS} approach is that the user spectrum can be allocated as a block, giving users flexibility of operating their signals and simplifying the management overhead of the hosting network operator. 
However, this limits the ability of the operator to identify and respond to faults or impairments generated by signals within such blocks. In this scenario, the operator is only 
able to monitor the overall power across that spectral window, without any further visibility on the number of channels allocated and individual channel power levels. 
We refer to this as user spectrum blind \ac{OSaaS}, differing it form the situation where the operator wavelength in the user spectral, which we call user spectrum aware configurations. 

In this work we focus on the spectrum blind configuration, where the network operators must make use of the correlation among the available data to assess impairments. In this case, the operator can only rely on coarse power information across the bandwidth and on the end-to-end performance of its own channels, which can be used as probes to indicate variations in performance. Thus, it is essential for the operator to develop tools for identifying impairments generated by \ac{OSaaS} users, even with incomplete or partial information about the underlying optical links. 

By targeting the impairment source, the operator can address both malicious and non-malicious user behaviors, maintaining service quality agreements with other users. Although the inter-channel interactions are complex, \ac{ML} can be leveraged to identify the misbehaving channels. While some recent studies have proposed \ac{ML} methods to identify failures from an end-user's standpoint~\cite{patri_machine_2023}, in this work we specifically address the challenge of identifying failures from the perspective of an operator providing \ac{OSaaS}. 

We carry out experiments on a 150km long \acf{OLS} with 6 uni-directional \acp{ROADM}, where one operator provides access to three \ac{OSaaS} users, each operating their own \ac{ROADM}. We propose that the operator can use its own data carrying channels as probes (i.e., acting as guard channels), allocated in between the \ac{OSaaS} windows of different users. We then introduce a flexible \ac{ML} model that utilizes data on these operator channels, along with coarse power readings at the in-line \acp{ROADM} to determine which user is causing the impairment. We achieve a classification accuracy rate of 94.2\%. 
\footnote{This paper is a preprint of a paper accepted to ECOC 2024 and is subject to Institution of Engineering and Technology Copyright. A copy of record will be available at IET Digital Library}
\vspace{-2mm}
\section{Experimental Setup}
\begin{figure}[t]
  \centering
  \includegraphics[width=1.0\linewidth]{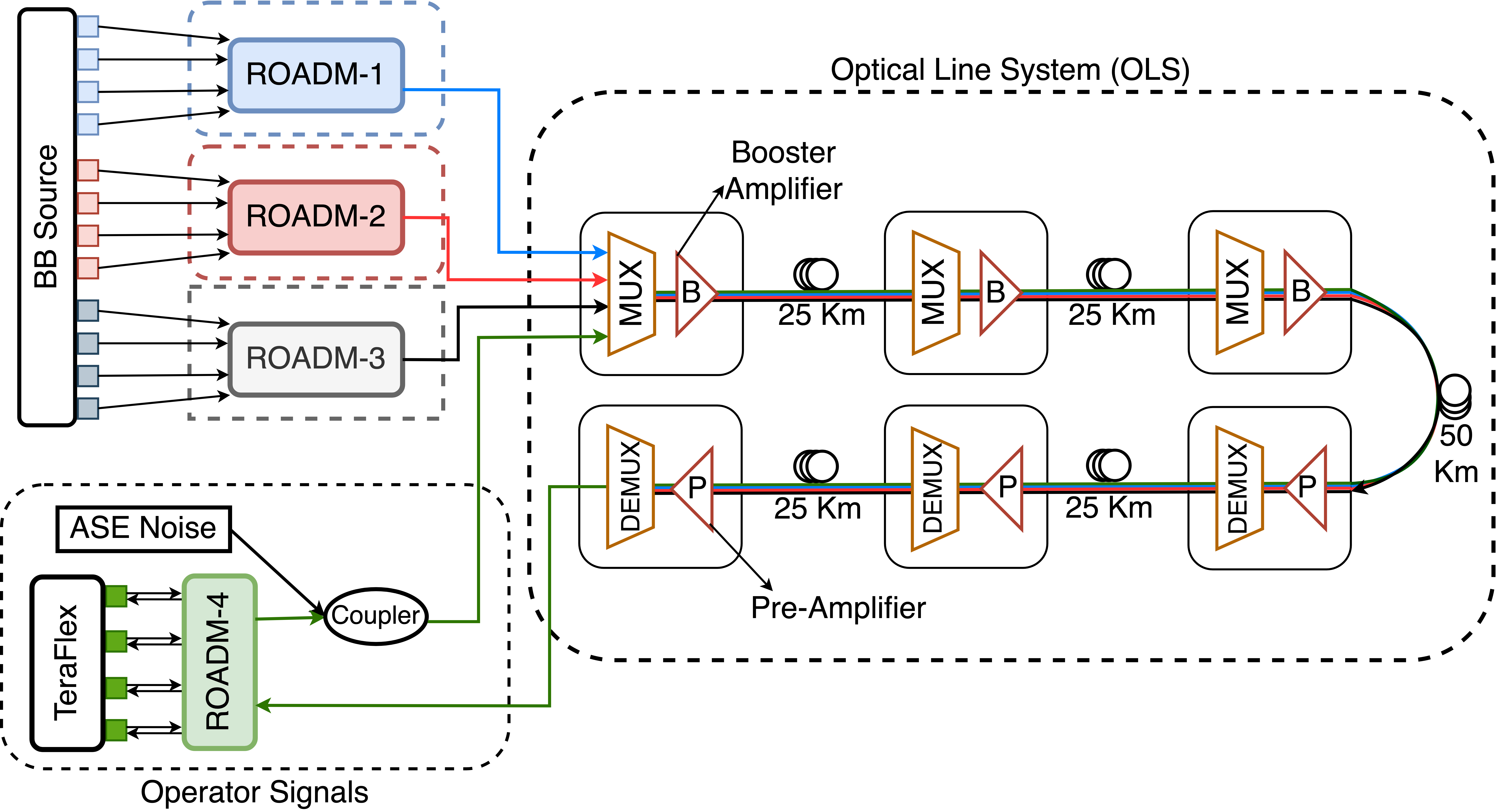}
\caption{Experimental setup in OpenIreland testbed.}
\vspace{-0mm}
\label{topology}
\end{figure}

\vspace{-0mm}
Our experimental setup is implemented in the OpenIreland Testbed~\cite{EDFA_ECOC, open_ireland}, and is shown in Fig.~\ref{topology}. The \ac{OLS} consists of three Lumentum \ac{ROADM}-20 units configured for uni-directional operation to represent 6 \ac{ROADM}s with \ac{WSS} filtering and amplification in each node. The \acp{ROADM} are connected by $4\times$25-km spans and one 50-km span. 4 Operator channels, acting as the data carrying probe/guard signals, are generated with ADVA Teraflex Transceivers, configured with 300-Gbit/s 64-QAM modulation. An auxiliary \ac{ROADM} multiplexed the operator channels, with added ASE noise, in order to collect performance data at 3 different \ac{OSNR} levels of operator probe channels.
\begin{figure}[t]
\vspace{-0mm}
  \centering
  \includegraphics[width=0.9\linewidth]{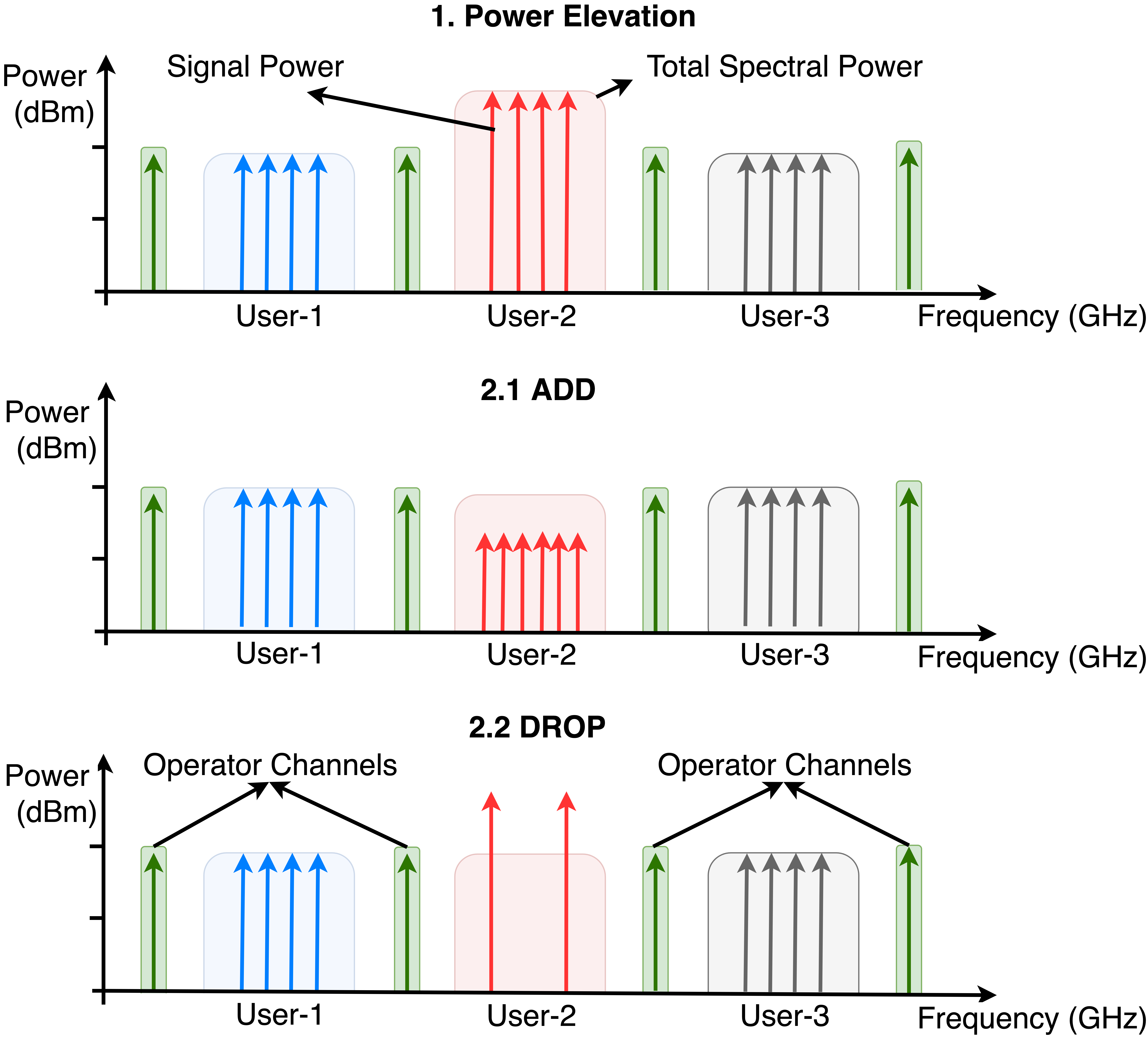}
\caption{Types of impairments induced: 1. Increase in power across OSaaS window, 2.1 Channel ADD impairment, 2.2 Channel DROP impairment}
\vspace{-3mm}
\label{impairments}
\end{figure}
We consider the scenario of a system with 3 \ac{OSaaS} end-users. Each user is allocated a bandwidth of 400-GHz across three non-overlapping network segments within the \ac{OLS}. These user channels were emulated by shaping an \ac{ASE} broadband source into 50 GHz channels with the edge \acp{ROADM}. At the setup phase, the \ac{OSaaS} end user channels are probed to ensure the \ac{QoT} is sufficient to enable signal transmission. 

Each user is allocated an edge \ac{ROADM} to configure channel loading and power levels, and multiplex the different user signals in one spectrum window. The signal from the operator, along with the user spectra from each individual user, are multiplexed together in the first \ac{ROADM} of the \ac{OLS}. The operator channels are inserted between different user spectra, and are shown in green color in Fig. \ref{impairments}. 
The launch power of each signal is set to 4 dBm at the input to each transmission span. The signals were equalized at the input \ac{ROADM} MUX, and Booster/Preamp amplifiers were set to a constant gain setting of 15 dB.
For each operator channel, we extracted 7 performance monitoring features from the Teraflex transceivers, namely \ac{CFO}, \ac{CDC}, \ac{DGD}, optical received power, \ac{OSNR}, Q-factor, \ac{PDL}, and the electrical \ac{SNR}. We also collected the power levels of the operator's channels in the \ac{OLS} through in-built \acp{OCM} in the \acp{ROADM}, and the total output power of each \ac{ROADM}. 
Data is collected from all network components through the \ac{NETCONF} protocol, polling the devices every 30 seconds. 
\vspace{-2mm}
\section{Impairments introduced by OSaaS users}

\begin{figure*}[ht]
  \centering
  \includegraphics[width=0.98\linewidth]{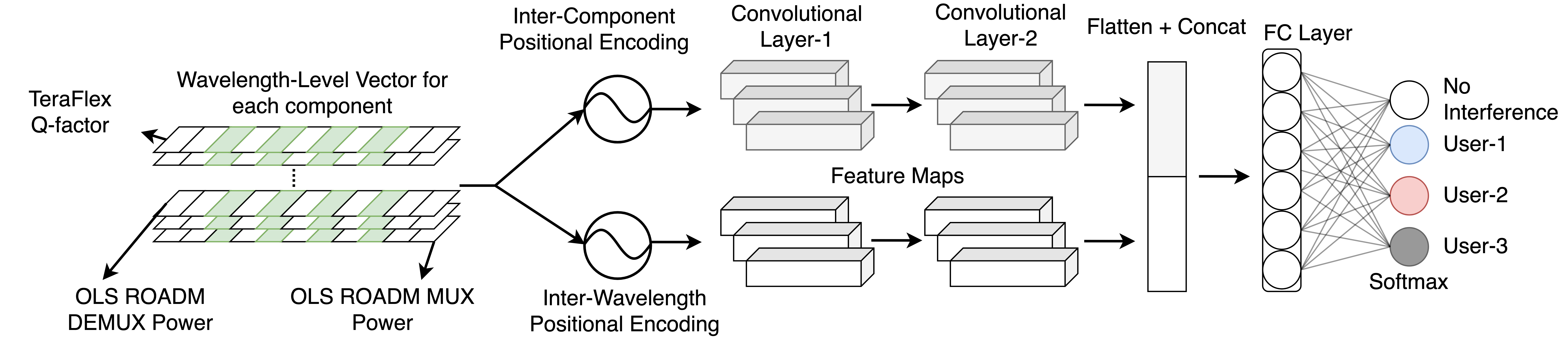}
  \caption{1D-CNN Model Architecture.}
  \label{model}
  \vspace{3mm}
\end{figure*}

\begin{figure*}[t]

\begin{minipage}[]{0.50\textwidth}
\centering
    \includegraphics[width=1.0\linewidth]{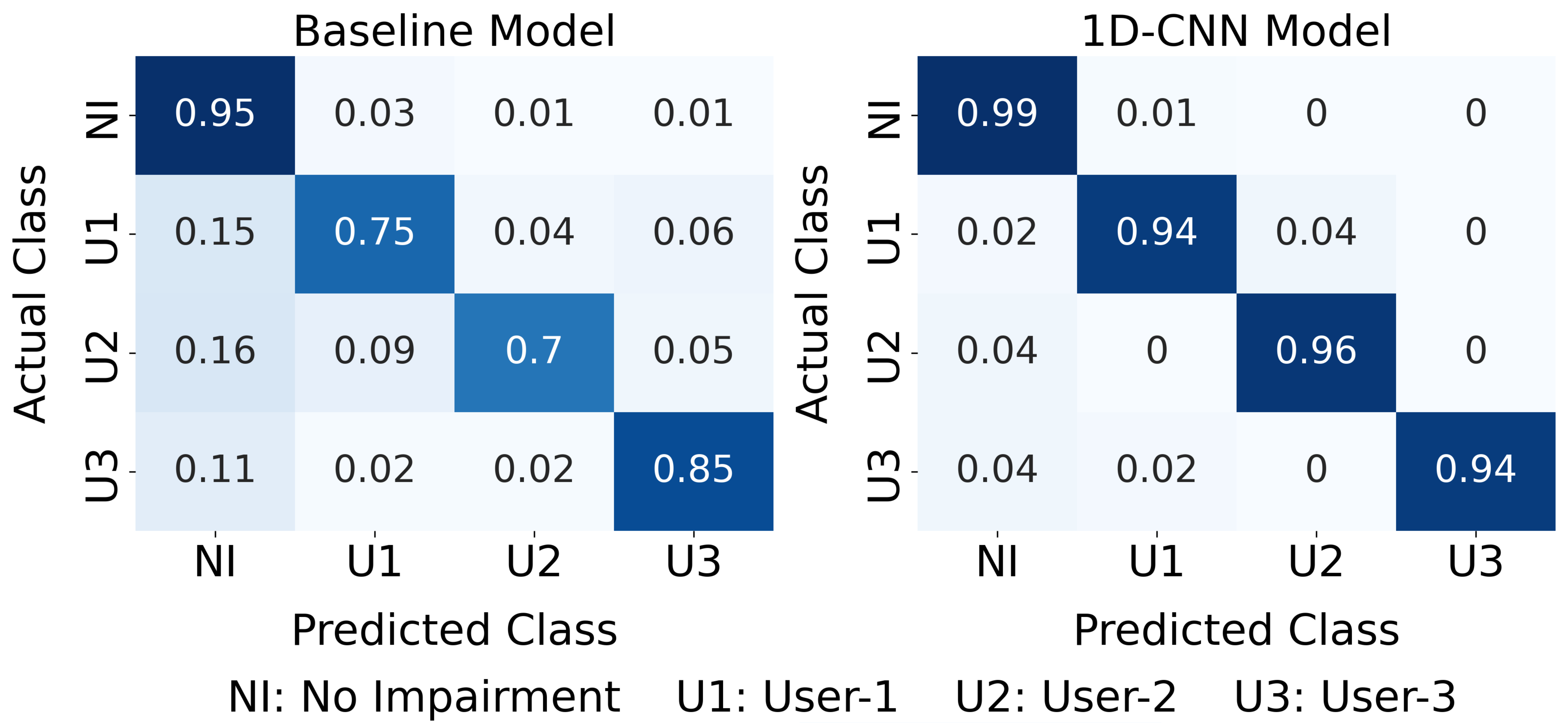}
    \caption{Confusion matrices of interferer classification for: baseline and 1D-CNN model. Values are normalized by sum of each row.}
    \label{confusion_matrix1}
    \vspace{2mm}
\end{minipage}%
\hspace{1mm}
\begin{minipage}{0.5\textwidth}
\centering
\fontsize{7.pt}{11pt}\selectfont
    \vspace{-20mm}\hspace{1cm}
\begin{tabular}{|c|c|c|c|c|c|c|}
    \hline
    \textbf{Impairment Type} & \textbf{User} & \multicolumn{2}{c|}{\textbf{Precision}} & \multicolumn{2}{c|}{\textbf{Recall}} \\
    \hline
    \multicolumn{2}{|c|}{} & \textbf{Base} & \textbf{CNN} & \textbf{Base} & \textbf{CNN}\\
    \hline
    \multirow{1}{*}{\textbf{No Impairment}} & & 97\% & 99\% & 95\% & 99\% \\
    \hline
    \multirow{3}{*}{\textbf{Power Increase}} & User-1 & 24\% & 83\% & 100\% & 100\%\\
    & User-2 & 56\% & 100\% & 83\% & 100\%\\
    & User-3 & 43\% & 85\% & 100\% & 100\%\\
    \hline
    \multirow{3}{*}{\textbf{ADD/DROP}} & User-1 & 62\% & 95\% & 72\% & 93\%\\
    & User-2 & 79\% & 94\% & 69\% & 94\%\\
    & User-3 & 71\% & 97\% & 82\% & 92\%\\
    \hline
\end{tabular}
    \caption{Classification Metrics}
    \vspace{-12mm}
    \label{table}
\end{minipage}
\vspace{-4mm}
\end{figure*}

To investigate the potential impairments from the \ac{OSaaS} users, we have focused on two main use cases, as shown in Fig. \ref{impairments}. We introduced these perturbations for each user into the \ac{OLS}, and measured the end-to-end performance of the operator's channels across three different \ac{OSNR} settings. 
We classify an observation as an impairment when the Q-factor of any of the operator channels falls below the \ac{OSNR} threshold, relative to the modulation format and baud rate used. In total, we have collected 2920 measurements, with 184 measurements for each user's impairment. 

\noindent \textbf{1. Increase in power in the OSaaS window}: The user's total spectrum power is systematically increased in 0.5 dB increments, across all channels, up to 6 dB.
This can generate impairments due to the EDFA cross-talk, caused by the non-flat spectral gain interacting with the \ac{AGC} mechanism, and in part also to the nonlinear crosstalk from nearby channels. 
This can in principle be detected by the operator, by reading the \ac{OCM} value of the entire \ac{OSaaS} window. However we focus our identification only on end-to-end performance monitoring of the operator's channels.

\noindent\textbf{2. ADD/DROP impairment:} 
We increase the power of the channels but reduce their number, so that the total power within the \ac{OSaaS} window remains constant. Higher channel power increases inter-channel non-linear interference~\cite{5420239}. These impairments are most difficult to identify, as the power levels across the spectral window remain constant, while the performance drops. These cannot be identified by the operator through its \ac{ROADM} \ac{OCM} for the user spectrum blind configuration considered here. 
\vspace{-2mm}
\section{Model Architecture}

Conventional \ac{ML} approaches for classification, such as Boosted Trees and \acp{ANN}, are not suitable to address this problem, as they possess static states, potentially limiting their adaptability to network expansions or new user additions. 
We thus propose and implement a novel architecture, based on a 1D-\ac{CNN} model, depicted in Fig. \ref{model}, tailored for dynamic networks. 
We incorporate sinusoidal positional encoding~\cite{vaswani_attention_2017} of dimensionality 2, which aims to capture positional information of channels and components. We implement positional encoding in two distinct contexts: wavelength and components. The wavelength encoding aims to identify dependencies like crosstalk based on spectral layout, while component encoding aims to track signal progression and interactions in the network. For each positional encoded data, two \ac{CNN} layers are applied. Each of the \ac{CNN} layers employ 3 distinct 1D kernels of size 5. The outputs of these \ac{CNN} layers are flattened and concatenated, then fed into a fully connected layer with 200 neurons. We apply \ac{ReLU} activation function after each layer of the model except the final output layer, which uses the softmax function to output the class probabilities. This layer outputs probabilities across 4 classes, to identify if a network state indicates an impairment and, if applicable, attributing the impairment to a specific user.

The model is trained for 1,200 epochs using the Adam Optimizer, minimizing cross-entropy error. We utilize a batch size of 32 and a learning rate of 0.001. Optimal parameters such as kernel size, neurons and activation functions were identified via a randomized-grid search. Data splitting for performance evaluation maintained a 3:1 training-to-test ratio, using stratified splits across classes, ensuring a balanced class distribution. 
\vspace{-1mm}
\section{Results}
We compare our model with a baseline 5 layer \ac{MLP} model, with 100 neurons each in the first 4 layers, and 4 neurons in the output layer for classifying impairments across the 4 classes.
In Fig.~\ref{confusion_matrix1}, the test set's confusion matrix is presented for the baseline \ac{MLP} model and the 1D-CNN model respectively. The 1D-CNN model achieves an overall user classification accuracy of 94.2\%, with a minimum of 94\% for each user. In comparison, the benchmark \ac{MLP} model achieves an overall user classification accuracy of 76.8\%, with many impairments remaining unidentified. 

Given the class imbalance, Tab.~\ref{table} provides precision, and recall scores for each user across both impairment types. Our model excels in identifying impairment sources causing overall power increase across the \ac{OSaaS} spectrum, even though identifying the specific \ac{OSaaS} user generating the impairing is non-trivial. Even for the second use case, where the total spectrum power remains constant, our model can identify the user causing the impairment with high accuracy (between 94\% and 96\% for different users). For impairment source identification, recall is a more relevant metric as the costs of a false negative is higher. Our model achieves a recall score greater than 90\% across all impairment types and users and performs much better than the \ac{MLP} model in the ADD/DROP impairment type. This shows the benefit of our model architecture, which takes into account the positions of user and operator spectra, as well as power evolution through the network. 
\vspace{-2mm}
\section{Conclusion}
We have examined multiple impairment scenarios in a multi-user \ac{OSaaS} network and introduced a methodology to embed operator's probe signals between the user spectra. 
Our results show that the model achieves a 94.2\% classification rate in predicting the interfering user, offering \ac{OSaaS} operators a potential tool to identify impairments and ensure service quality. 

\clearpage
\section{Acknowledgements}
This work was supported by Science Foundation Ireland (SFI) under grants 12/RC/2276 p2, 22/FFP-A/10598, 18/RI/5721, and 13/RC/2077 p2, and by National Science Foundation (NSF) under grants CNS-1827923, OAC-2029295, CNS-2112562, CNS-2211944, and CNS-2330333.

\printbibliography

\vspace{-4mm}

\end{document}